\documentclass[11pt]{article}
\usepackage{jheppub}
\usepackage{amsmath}
\usepackage{amssymb}
\usepackage{braket}
\usepackage{bbold}
\usepackage{pifont}
\usepackage{color}

\newcommand{\beqn}{\begin{eqnarray}}
\newcommand{\eeqn}{\end{eqnarray}}
\newcommand{\dd}{\mathrm{d}}
\newcommand{\nn}{\nonumber}

\newcommand{\fmn}{f_{\mu\nu}}

\newcommand{\te}{{\tilde{e}}}
\newcommand{\be}{{\bar{e}}}			
\newcommand{\tE}{{\varphi}}	
\newcommand{\EE}{\mathcal{E}}	
\newcommand{\gmn}{g_{\mu\nu}}

\newcommand{\bmn}{B_{\mu\nu}}

\title{Vierbein interactions with antisymmetric components}
\author{Chrysoula Markou, Felix J. Rudolph \& Angnis~Schmidt-May}
\affiliation{Max-Planck-Institut f\"ur Physik (Werner-Heisenberg-Institut)\\
F\"ohringer Ring 6, 80805 Munich, Germany}
\emailAdd{cmarkou@mpp.mpg.de, frudolph@mpp.mpg.de, angnissm@mpp.mpg.de}

\abstract{In this work we propose a new gravitational setup formulated in terms of two interacting
vierbein fields. The theory is the fully diffeomorphism and local Lorentz invariant extension 
of a previous construction which involved a fixed reference vierbein. 
Certain vierbein components can be shifted by local Lorentz transformations 
and do not enter the associated metric tensors. We parameterize these components by an antisymmetric tensor
field and give them a kinetic term in the action, thereby promoting them to dynamical variables. 
In addition, the action contains two Einstein-Hilbert terms and an interaction potential
whose form is inspired by ghost-free massive gravity and bimetric theory.
The resulting theory describes the interactions of a massless spin-2, a massive spin-2 and an 
antisymmetric tensor field.
It can be generalized to the case of multiple massive spin-2 fields and multiple antisymmetric tensor fields.
The absence of additional and potentially pathological degrees of freedom is verified in an ADM analysis.
However, the antisymmetric tensor fluctuation around the maximally symmetric background solution
has a tachyonic mass pole.}

\begin{document} 

 \begin{flushright}
 \hfill{MPP-2018-297} \vspace{20mm}
 \end{flushright}
\maketitle
\flushbottom

\section{Introduction}

The construction of classically consistent field theories is an on-going challenge
which becomes disproportionately more complicated with increasing spin of the involved fields. 
The set of consistent interactions for fields up to spin-1 in flat space is reasonably 
well-understood but, when including gravity, the non-linearities of gravitational interactions 
and the spin-2 nature of the gravitational field always 
introduce further complexity. For a recent review on the programme of building new field theories
in the presence of gravity, see Ref.~\cite{Heisenberg:2018vsk}.

A well-known example for a general class of consistent field theories 
including gravity is the Horndeski action, which contains the most general scalar (i.e.~spin-0)
interactions with second-order equations of motion~\cite{Horndeski:1974wa}.
The action has been generalized to the ``beyond Horndeski"  class,
which, despite its higher-order equations, does not give rise to Ostrogradski instabilities that would
threaten the theory's consistency~\cite{Gleyzes:2014dya, Gleyzes:2014qga}.
A recent further generalization of the consistent setup is the so-called
``DHOST" (degenerate higher-order scalar-tensor) theory~\cite{Langlois:2015cwa, BenAchour:2016fzp}.
Examples for nonlinear vector (i.e.~spin-1) interactions whose particular structure is chosen to 
avoid instabilities are standard Yang-Mills actions for massless fields~\cite{Yang:1954ek},
and the more recently constructed generalized Proca actions for a self-interacting massive
field~\cite{Tasinato:2014eka, Heisenberg:2014rta}.

In the spin-2 case, the Einstein-Hilbert action for general relativity (GR) delivers the 
nonlinear self-interactions for the massless field. The mass term that can be added to this 
theory has a very particular structure which is fixed by the absence of the Boulware-Deser
ghost instability~\cite{Boulware:1973my}. The nonlinear theory for massive gravity was 
constructed and shown to be ghost-free only a few years 
ago~\cite{deRham:2010kj, Hassan:2011hr, Hassan:2011tf}. 
Formulating the action requires the introduction of a reference or
fiducial metric tensor. This second metric can be promoted to a dynamical field, resulting
in a classically consistent bimetric theory, which describes the nonlinear interactions
of a massless and a massive spin-2 field~\cite{Hassan:2011zd}. 
The setup of consistent spin-2 interactions can also be generalized to the case of 
multiple massive fields~\cite{Hinterbichler:2012cn}.
For reviews on massive gravity and bimetric theory, see Ref.~\cite{Hinterbichler:2011tt, deRham:2014zqa} 
and \cite{Schmidt-May:2015vnx}, respectively.

The structure of the allowed massive spin-2 interactions assumes a remarkably simple form
when written in terms of the vierbein fields related to the two metrics~\cite{Hinterbichler:2012cn}.
Moreover, the most general set of consistent interactions among multiple massive spin-2 fields
actually requires the vierbein formulation~\cite{Hassan:2018mcw}. This is not particularly 
astonishing since there is an example for this situation already in GR, 
where couplings to fermionic fields can only be expressed using the
vierbein field. 

The vierbein $e^a_{~\mu}$, related to the corresponding metric via 
$\gmn=e^a_{~\mu}\eta_{ab}e^b_{~\nu}$, has 16 independent components. Only 10 of these show up
in the metric tensor since the latter is invariant under local Lorentz transformations,
$e^a_{~\mu}\mapsto \Lambda^a_{~c}e^c_{~\mu}$ with $\Lambda^a_{~c}\eta_{ab}\Lambda^b_{~d}=\eta_{cd}$.
In GR, the remaining 6 components are pure gauge and hence unphysical. Massive gravity and 
bimetric theory in vierbein formulation a priori contain these Lorentz components 
as dynamical fields, but they are required to vanish by the equations of motion.
This property ensures the existence of an equivalent formulation in terms of metric
tensors~\cite{Deffayet:2012zc}. Moreover, as was shown in Ref.~\cite{deRham:2015cha}, the
vanishing of the 6 additional Lorentz components is crucial for the consistency of the theory.
It is also related to the possibility of having causal propagation~\cite{Hassan:2017ugh}.\footnote{In
this paper we will not address the issue of causality. The term ``consistent"
refers to the absence of ghost instabilities. For work on causal propagation in 
massive spin-2 theories in the framework of scattering amplitudes see Ref.~\cite{Hinterbichler:2017qyt,
Bonifacio:2017nnt,deRham:2017zjm,deRham:2017xox,deRham:2018qqo}.}

Interestingly, the consistency problem of non-vanishing Lorentz components can be overcome
by giving them a kinetic term.\footnote{The idea of making the Lorentz components dynamical
in massive gravity was first mentioned in Ref.~\cite{Gabadadze:2013ria}. At the linearized level, 
their dynamics are discussed in the context of teleparallel theories 
in section 4.6 of~\cite{Ortin:2004ms}.}
This was demonstrated explicitly in Ref.~\cite{Markou:2018wxo}
for the case of a fixed reference frame field $\te^a_{~\mu}$ (i.e.~for the massive gravity 
case). The resulting theory describes the nonlinear interactions of a massive spin-2
with a massive antisymmetric tensor field. No additional degrees of freedom, which could
potentially give rise to instabilities, enter at the nonlinear level. 
The structure of the resulting action is interesting because it involves the antisymmetric
tensor combination $\bmn=e^a_{~\mu}\eta_{ab}\te^b_{~\nu}-\te^a_{~\mu}\eta_{ab}e^b_{~\nu}$.
The mass pole of its fluctuation around maximally symmetric
backgrounds is tachyonic, indicating the instability of this vacuum solution.

Antisymmetric tensor fields, first considered in Ref.~\cite{Ogievetsky:1967ij, Kalb:1974yc},
are objects of interests in supergravity theories, 
and thus in low-energy effective descriptions of string theory. 
Together with the graviton and the dilaton, they make up the massless bosonic 
excitations of the string. 
In $D=4$ spacetime dimensions, antisymmetric tensors are dual to a scalar in the massless 
and to a vector in the massive case (see e.g.~\cite{Smailagic:2001ch}), but in higher 
dimensions the duality does not necessarily relate them to lower-spin fields.

In this work we will build on the results of Ref.~\cite{Markou:2018wxo} and extend the setup by
giving dynamics to the frame field $\te^a_{~\mu}$.

\subsection*{Summary of results}

We demonstrate that the action proposed in Ref.~\cite{Markou:2018wxo} can be 
generalized to a fully dynamical theory for two interacting vierbein fields
$e^a_{~\mu}$ and $\te^a_{~\mu}$. The result is a ghost-free bimetric action
in vierbein formulation with dynamical Lorentz components which we
parameterize in terms of the antisymmetric components
$\bmn=e^a_{~\mu}\eta_{ab}\te^b_{~\nu}-\te^a_{~\mu}\eta_{ab}e^b_{~\nu}$.
The action is manifestly invariant under local Lorentz transformations and diffeomorphisms.
The number of propagating degrees of freedom in this new setup is $2+5+3$,
corresponding to a massless spin-2, a massive spin-2 and a massive antisymmetric
tensor. This is verified both at the linear and at the fully nonlinear level.
The mass of the antisymmetric fluctuation around the maximally symmetric background 
is again tachyonic, implying that the bimetric vacuum of the extended theory is unstable.
We then show how to further generalize the setup to the case of $\mathcal{N}$ interacting
vierbeine and their independent antisymmetric components which are packaged 
into $(\mathcal{N}-1)$ antisymmetric tensor fields.

\subsection*{Conventions}

We work with metric signature $(-,+,+,+)$ in 4 spacetime dimensions for definiteness, 
but all of our results generalize to arbitrary dimension. 
Spacetime indices are denoted by Greek letters $\mu,\nu$, Lorentz indices 
by Latin letters $a,b$. Indices are raised and lowered by $\gmn$ 
and the inverse $g^{\mu\nu}$ on its curvatures
and on objects related to the antisymmetric tensor. Indices on curvatures of $\fmn$
are raised and lowered with $\fmn$ and its inverse $f^{\mu\nu}$. 
Lorentz indices are raised and lowered with $\eta_{ab}$ and 
its inverse $\eta^{ab}$. Brackets denoting symmetrization and antisymmetrization of 
indices are defined as $T_{\mu\nu} = T_{(\mu\nu)} + T_{[\mu\nu]}$ with 
$T_{(\mu\nu)} = \frac12(T_{\mu\nu} + T_{\nu\mu})$ and $T_{[\mu\nu]} 
= \frac12(T_{\mu\nu} - T_{\nu\mu})$.

\newpage

\section{Review of gravity with antisymmetric components}\label{sec:review}

Here we briefly review the results of Ref.~\cite{Markou:2018wxo}, discussing first
the massless case before adding the mass term.

\subsection{Massless fields}

The action for a massless antisymmetric tensor field $\bmn$ minimally coupled to a 
massless metric $\gmn$ is,
\begin{equation}\label{gbaction}
S_\mathrm{gB} = m_g^2\int\dd^4x ~\sqrt{g} ~[R(g) - 2\Lambda] 
-  \frac{m_B^2}{2\cdot3!}\int\dd^4x~ \sqrt{g}~H_{\mu\nu\rho}H^{\mu\nu\rho}\,,
\end{equation}
where $H_{\mu\nu\rho}=3 \nabla_{[\mu}B_{\nu\rho]}$ is the 2-form field strength.
We have included a Planck mass $m_g$ for $\gmn$ and also a mass scale $m_B$ for $\bmn$,
such that both tensor fields are dimensionless.\footnote{Note that our convention here 
slightly differs from the one in Ref.~\cite{Markou:2018wxo}, where $\bmn$ had mass 
dimension 1.}

In Ref.~\cite{Markou:2018wxo}, we repackaged the 10+6 components contained in the
tensor fields into the vierbein $e^a_{~\mu}$. This was achieved by making the following
identifications,
\begin{subequations}
\beqn
g_{\mu\nu} &\equiv& {e_\mu}^a\eta_{ab}{e^b}_\nu \,,\label{gdef}\\
\bmn&\equiv& e^a_{~\mu}\eta_{ab}\te^b_{~\nu}- \te^a_{~\mu}\eta_{ab}e^b_{~\nu}\,,\label{bdef}
\eeqn
\end{subequations}
where the auxiliary vierbein $\te^a_{~\mu}$ defines a fixed reference frame. 
For instance, one could take $\te^a_{~\mu}=\delta^a_{~\mu}$. 
The symmetric field $\gmn$ is the spacetime metric with ordinary relation to the dynamical tetrad.
It is invariant under local Lorentz transformations $e^a_{~\mu}\mapsto \Lambda^a_{~b}e^b_{~\mu}$
with $\Lambda^a_{~b}\eta_{ac}\Lambda^c_{~d}=\eta_{bd}$ and therefore depends on only 10 of the 16
components in $e^a_{~\mu}$. The remaining 6 components enter the antisymmetric tensor $\bmn$.

The equations of motion for the vierbein following from the above action read,
\beqn
\EE_a^{~\mu} \equiv\frac{\delta S_\mathrm{gB}}{\delta e^a_{~\mu}} 
	= 2\eta_{ab}{e^b}_\nu\mathcal{G}^{\mu\nu} + 2\eta_{ab} \te^b{}_\nu\mathcal{B}^{\mu\nu}
	=0\,.
\eeqn
Here we have defined,
\begin{subequations}\label{gbeq}
\begin{align}
\mathcal{G}^{\mu\nu}=\mathcal{G}^{\nu\mu} &\equiv   R^{\mu\nu} -\frac12(R-2\Lambda)g^{\mu\nu} 
	- \frac{m_B^2}{4m_g^2}(H^{\mu\rho\sigma}H^\nu{}_{\rho\sigma} - \frac16 H^2 g^{\mu\nu}) \,, \\
\mathcal{B}^{\mu\nu}=-\mathcal{B}^{\nu\mu} &\equiv -\frac{m_B^2}{2m_g^2}\nabla_\rho H^{\rho\mu\nu} \, ,
\end{align}
\end{subequations}
which correspond to the variations of the action with respect to the tensor fields.
In the tensor formulation, $\mathcal{G}^{\mu\nu}$ and $\mathcal{B}^{\mu\nu}$ vanish separately.
In fact, this is also the case in the vierbein formulation, as can be seen by looking at the 
antisymmetric combination of equations,
$
2\eta^{ab} e^{[\mu}{}_a  \EE^{\nu]}_b = 0 
$,
which implies $\mathcal{B}^{\mu\nu}=0$. Hence the vierbein and 
tensor formulations of the massless theory are equivalent.

\subsection{Massive fields}

Inspired by ghost-free massive gravity~\cite{deRham:2010kj, Hassan:2011hr, Hassan:2011tf, Hinterbichler:2012cn}, 
Ref.~\cite{Markou:2018wxo} added the following interaction term $S_V$ for the vierbein 
$e^a_{~\mu}$ to the massless action,
\beqn\label{defpot3}
&-&m_g^2m^2\int \epsilon_{abcd}
\Big(
b_1 ~e^a\wedge e^b \wedge e^c \wedge \te^d
+b_2 ~e^a\wedge e^b \wedge \te^c \wedge \te^d
+ ~b_3~ e^a\wedge \te^b \wedge \te^c \wedge \te^d
\Big)\,.
\eeqn
It was shown in a nonlinear ADM analysis that these interactions make both the fields $\gmn$
and $\bmn$ massive without introducing additional degrees of freedom. The action with
this mass term propagates 5+3 degrees of freedom, corresponding to a massive spin-2 and
a massive antisymmetric tensor (which is dual to a massive vector in $D=4$).
The latter is a tachyon at the linearized level.

Denoting the variation of the mass term by
$\mathcal{V}_a^{~\mu}\equiv -\frac{1}{m_g^2\det e}\frac{\delta S_V}{\delta e^a_{~\mu}}$,
the vierbein equations of motion now assume the form, 
\begin{equation}\label{fulleq}
\EE_a^{~\mu} = 2\eta_{ab}{e^b}_\nu\mathcal{G}^{\mu\nu} + 2\eta_{ab} \te^b{}_\nu\mathcal{B}^{\mu\nu} 
+ \mathcal{V}_a^{~\mu}=0\, .
\end{equation}
These equations can still be separated into a symmetric and an antisymmetric part which read,
\begin{subequations}
\beqn
\mathcal{B}^{\mu\nu}- (\tilde{P}^{-1})_{\rho\sigma}{}^{ab}\te^{\mu}_{~a}\te^{\nu}_{~b}
e^{\rho}_{~c}\eta^{cd}\mathcal{V}^{~\sigma}_d&=&0 \,,\\
\label{massbeq}
\mathcal{G}^{\mu\nu} +  
(\tilde{P}^{-1})_{\rho\sigma}{}^{ab}\te^{\mu}_{~a}e^{\nu}_{~b}
e^{\rho}_{~c}\eta^{cd}\mathcal{V}^{~\sigma}_d
        + \frac12 e^\mu_{~a}\eta^{ab}\mathcal{V}^{~\nu}_b &=& 0 \, .
        \label{massgeq}
\eeqn 
\end{subequations}
Here, $\tilde{P}^{-1}$ denotes the inverse of the operator 
$\tilde{P}^{\mu\nu}{}_{ab}\equiv 2e^{[\mu}{}_{[a}\te^{\nu]}{}_{b]}$
which is invertible on the space of antisymmetric matrices.

\section{Dynamical reference frame}\label{sec:dynte}

The vierbein action with potential \eqref{defpot3} explicitly breaks diffeomorphism
and local Lorentz invariance, due to the presence of the fixed 
reference frame $\te^a_{~\mu}$. For various reasons it is desirable to restore 
these symmetries, which can be achieved by introducing dynamics for the reference 
vierbein $\te^a_{~\mu}$, as we shall do in the following.

\subsection{Action and equations}
The reference vierbein defines a second metric tensor,
\beqn\label{fdef}
f_{\mu\nu} \equiv {\te_\mu}^{~a}\eta_{ab}{\te^b}_{~\nu} \,.
\eeqn 
We can make it dynamical by augmenting the massive action by an Einstein-Hilbert term 
for $\fmn$. The full theory thus reads,
\beqn\label{bgbaction}
S_\mathrm{m}
&=& m_g^2\int\dd^4x \,\sqrt{g} \,\Big(R(g) -2\Lambda\Big)
+m_f^2\int\dd^4x \,\sqrt{f} \,\Big(R(f) -2\tilde{\Lambda}\Big)
-  \frac{m_B^2}{2\cdot3!}\int\dd^4x~ \sqrt{g}~H_{\mu\nu\rho}H^{\mu\nu\rho}
\nn\\
&-&m_g^2m^2\int \epsilon_{abcd}
\Big(
b_1 ~e^a\wedge e^b \wedge e^c \wedge \te^d
+b_2 ~e^a\wedge e^b \wedge \te^c \wedge \te^d
+ ~b_3~ e^a\wedge \te^b \wedge \te^c \wedge \te^d
\Big)\,.
\eeqn
For $\bmn=0$, it reduces to ghost-free bimetric theory in vierbein 
formulation~\cite{Hassan:2011zd, Hinterbichler:2012cn}.
For $\bmn\neq 0$ it is not obvious that the kinetic term for the antisymmetric
components does not re-introduce the Boulware-Deser ghost. It is not obvious either
that the dynamics for $\te^a_{~\mu}$ do not destroy the consistency of the model with
fixed reference frame. In appendix~\ref{app:ADM} we perform a 3+1 split of the fields
and explicitly show that the number of propagating degrees of freedom is $2+5+3$, corresponding
to a massless spin-2, a massive spin-2 and a massive antisymmetric field. 
The Boulware-Deser ghost is removed by a constraint, just like in ghost-free bimetric
theory. 

The action could in principle contain other ghosts, hidden in the kinetic terms for
$\gmn$, $\fmn$ and $\bmn$. In the following we show that the equations of motion 
can again be separated in a way that preserves the kinetic structures with respect 
to the massless theory. This implies the absence of kinetic mixing introduced by the
mass term, which is promising for the consistency of the theory.

Defining $\tilde{\mathcal{V}}_a^{~\mu}\equiv -\frac{1}{m_f^2\det \te}\frac{\delta S_V}{\delta \te^a_{~\mu}}$
and using again
$\mathcal{V}_a^{~\mu}= -\frac{1}{m_g^2\det e}\frac{\delta S_V}{\delta e^a_{~\mu}}$,
the equations of motions for $e^a_{~\mu}$ and $\te^a_{~\mu}$, respectively, read,
\begin{subequations}\label{vbeq}
\beqn
\EE_a^{~\mu} 
&=& 2\eta_{ab}{e^b}_\nu\mathcal{G}^{\mu\nu} + 2\eta_{ab} \te^b{}_\nu\mathcal{B}^{\mu\nu} 
+ \mathcal{V}_a^{~\mu}=0\, ,
\\
\tilde{\EE}_a^{~\mu} 
&=& 2\eta_{ab}{\te^b}_{~\nu}\mathcal{F}^{\mu\nu} - \frac{1}{\alpha^2}\frac{\det e}{\det \te}\,
2\eta_{ab} e^b{}_\nu\mathcal{B}^{\mu\nu} 
+\tilde{\mathcal{V}}_a^{~\mu}=0\, .
\eeqn
\end{subequations}
Here, in addition to \eqref{gbeq} we have used the definitions,\footnote{On 
curvatures of the metric $\fmn$, we raise indices with the inverse metric $f^{\mu\nu}$.}
\beqn
\mathcal{F}^{\mu\nu} = 
\mathcal{F}^{\nu\mu} \equiv R^{\mu\nu}(f) -\frac12\Big(R(f)-2\tilde{\Lambda}\Big)f^{\mu\nu}\,,
\eeqn
and,
\beqn
\alpha\equiv \frac{m_f}{m_g}\,.
\eeqn
Using exactly the same arguments as in the case with non-dynamical $\te^a_{~\mu}$,
it is easy to show that either of the antisymmetric combinations of equations,
\begin{equation}\label{antsymmeq}
2\eta^{ab} e^{[\mu}{}_a  \EE^{\nu]}_b = 0 \,,
\qquad
2\eta^{ab} \te^{[\mu}{}_a  \tilde{\EE}^{\nu]}_b = 0\,,
\end{equation}
implies,
\beqn
\mathcal{B}^{\mu\nu}- (\tilde{P}^{-1})_{\rho\sigma}{}^{ab}\te^{\mu}_{~a}\te^{\nu}_{~b}
e^{\rho}_{~c}\eta^{cd}\mathcal{V}^{~\sigma}_d&=&0 \,.
\eeqn
The fact that the two equations in (\ref{antsymmeq}) are equivalent is  
a direct consequence of the invariance of the action under diagonal local Lorentz 
transformations, which we shall discuss below.
Plugging the expressions for $\mathcal{B}^{\mu\nu}$ back into the full equations,
we obtain, 
\begin{subequations}\label{symeq}
\beqn
\mathcal{G}^{\mu\nu} +
(\tilde{P}^{-1})_{\rho\sigma}{}^{ab}\te^{\mu}_{~a}e^{\nu}_{~b}
e^{\rho}_{~c}\eta^{cd}\mathcal{V}^{~\sigma}_d
        + \frac12 e^\mu_{~a}\eta^{ab}\mathcal{V}^{~\nu}_b &=& 0 \, ,\\
\mathcal{F}^{\mu\nu} + (\tilde{P}^{-1})_{\rho\sigma}{}^{ab}e^{\mu}_{~a}\te^{\nu}_{~b}
\te^{\rho}_{~c}\eta^{cd}\tilde{\mathcal{V}}^{~\sigma}_d
        + \frac12 \te^\mu_{~a}\eta^{ab}\tilde{\mathcal{V}}^{~\nu}_b
 &=& 0\,,
\eeqn 
\end{subequations}
where $\tilde{P}^{\mu\nu}{}_{ab}\equiv 2e^{[\mu}{}_{[a}\te^{\nu]}{}_{b]}$
is the same invertible operator as before.

The vierbein equations in \eqref{vbeq} thus separate into one set of antisymmetric components,
corresponding to either of the two equivalent expressions in (\ref{antsymmeq}),
and two sets of symmetric components in \eqref{symeq}. 
The kinetic structures are exactly those of the massless theory.

\subsection{Local symmetries}

The action in \eqref{bgbaction} is invariant under the following symmetry 
transformations.
\begin{itemize}

\item Local Lorentz transformations which infinitesimally transform the vierbeine as,
\beqn
\Delta_\omega e^a_{~\mu}=\eta^{ab}\omega_{bc}e^c_{~\mu}\,,
\qquad 
\Delta_\omega \te^a_{~\mu}=\eta^{ab}\omega_{bc}\te^c_{~\mu}\,,
\eeqn
with $\omega_{bc}=-\omega_{cb}$. The transformation is diagonal since
the gauge parameters $\omega_{bc}$ are the same for both fields.
This is an obvious symmetry: In both 
metrics as well as in the antisymmetric tensor all Lorentz indices 
are contracted with the invariant tensor $\eta_{ab}$ while in the interaction
potential they are contracted with the invariant tensor $\epsilon_{abcd}$. 

\item Diffeomorphisms which infinitesimally transform the vierbeine as,
\beqn\label{vbdiff}
\Delta_\xi e^a_{~\mu}= \xi^\rho\nabla_\rho e^a_{~\mu}+e^a_{~\rho}\nabla_\mu\xi^\rho\,,
\qquad 
\Delta_\xi \te^a_{~\mu}= 
\xi^\rho\tilde{\nabla}_\rho \te^a_{~\mu}+ \te^a_{~\rho}\tilde{\nabla}_\mu\xi^\rho\,.  
\eeqn
These transformations correspond to the diagonal subgroup of the diff\,$\times$\,diff symmetry
which is broken by the mass term and the kinetic term for $\bmn$.
In fact the covariant derivatives in the transformations can be taken to be with respect 
to either metric since the Christoffel symbols of the two terms cancel each other out,
\beqn
\Delta_\xi \te^a_{~\mu}= 
\xi^\rho\partial_\rho \te^a_{~\mu}+ \te^a_{~\rho}\partial_\mu\xi^\rho
-\xi^\rho\tilde{\Gamma}^\sigma_{\rho\mu}\te^a_{~\sigma} + \te^a_{~\rho}\tilde{\Gamma}^\rho_{\mu\sigma}\xi^\sigma
=\xi^\rho\partial_\rho \te^a_{~\mu}+ \te^a_{~\rho}\partial_\mu\xi^\rho\,,
\eeqn
where we have used $\tilde{\Gamma}^\sigma_{\rho\mu}=\tilde{\Gamma}^\sigma_{\mu\rho}$. Hence we
can also write the transformation of $\te^a_{~\mu}$ as,
\beqn
\Delta_\xi \te^a_{~\mu}= 
\xi^\rho{\nabla}_\rho \te^a_{~\mu}+ \te^a_{~\rho}{\nabla}_\mu\xi^\rho\,,  
\eeqn
which is the proper transformation of a vector under diffeomorphisms of the metric 
$\gmn$ compatible with $\nabla$.
It then follows that the combination $e^{~a}_{\mu}\eta_{ab}\te^b_{~\nu}$ as well as its
symmetric and antisymmetric parts transform as tensors under the diagonal diffeomorphisms.
Thus we have the desired transformation property of $\bmn$,
\beqn
\Delta_\xi \bmn=
\xi^\rho\nabla_\rho \bmn
+B_{\mu\rho}\nabla_\nu\xi^\rho
+B_{\rho\nu}\nabla_\mu\xi^\rho\,,
\eeqn
which can of course also be verified explicitly using \eqref{vbdiff}. The action is 
therefore invariant under the diagonal diffeomorphism transformations of the vierbeine.
\end{itemize}

\subsection{Linear theory}
We will now derive the spectrum of linear perturbations around maximally symmetric backgrounds.
These solutions are obtained by making the 
ansatz $\te^a_{~\mu}=ce^a_{~\mu}$, for which the equations reduce to,
\beqn
\bmn=0\,,\qquad
R_{\mu\nu}(g)=\Lambda_g\gmn\,,\qquad
R_{\mu\nu}(c^2g)=\Lambda_f\gmn\,.
\eeqn
Here we have defined the background curvatures,
\beqn
\Lambda_g&=& \Lambda+3m^2\Big(  3  b_1c  + 2 b_2c^2 +  b_3c^3 \Big)\,,\nn\\
\Lambda_f&=&c^2\tilde{\Lambda}
+\frac{3m^2}{\alpha^2c^2}\Big(    b_1c  + 2 b_2c^2 +  3b_3c^3\Big)\,.
\eeqn
Since $R_{\mu\nu}(g)=R_{\mu\nu}(c^2g)$, we obtain the background condition,
\beqn
\Lambda_g=\Lambda_f\,,
\eeqn
which is a polynomial equation in $c$ whose roots fully determine the background solution.
Next, we consider linear perturbations around the proportional backgrounds,
\beqn
e^a_{~\mu}=\bar{e}^a_{~\mu}+\delta e^a_{~\mu}\,,\qquad
\te^a_{~\mu}=c\bar{e}^a_{~\mu}+\delta \te^a_{~\mu}\,.
\eeqn
These can be combined into the three linear fluctuations of the tensor fields,
\beqn
\delta \gmn &\equiv& \gmn-\bar{g}_{\mu\nu}= 2\delta e^a_{~(\mu}\be^b_{~\nu)}\eta_{ab}, \qquad
\delta \fmn\equiv \fmn-\bar{f}_{\mu\nu} = 2c\delta \te^a_{~(\mu}\be^b_{~\nu)}\eta_{ab}, \nn\\
\delta \bmn &=& 2\big(c\delta e^a_{~[\mu}\be^b_{~\nu]} -  \delta \te^a_{~[\mu}\be^b_{~\nu]} \big)\eta_{ab} 
\eeqn
It is then straightforward to show that the linearized equations of motions 
can be diagonalized into the following three equations,
\begin{subequations}\label{lineq}
\beqn
\mathcal{E}_{\mu\nu}^{~~\rho\sigma}m_{\rho\sigma}
-\Lambda_g\big(m_{\mu\nu}-\tfrac{1}{2}m_{\rho\sigma}\bar{g}^{\rho\sigma}\bar{g}_{\mu\nu}\big)
-\tfrac{m_\mathrm{FP}^2}{2}\big(m_{\mu\nu}-m_{\rho\sigma}\bar{g}^{\rho\sigma}\bar{g}_{\mu\nu}\big)&=&0\,,
\\
\mathcal{E}_{\mu\nu}^{~~\rho\sigma}l_{\rho\sigma}
-\Lambda_g\big(l_{\mu\nu}-\tfrac{1}{2}l_{\rho\sigma}\bar{g}^{\rho\sigma}\bar{g}_{\mu\nu}\big)
&=&0\,,\\
\bar{\nabla}^\rho\bar{\nabla}_{[\rho} b_{\mu\nu]}-m_b^2b_{\mu\nu}&=&0\,,
\eeqn
\end{subequations}
where we have defined,
\beqn
m_{\mu\nu}\equiv \delta \gmn - \frac1{c^2}\delta \fmn  \,,\qquad
l_{\mu\nu}\equiv \delta \gmn + \alpha^2 \delta \fmn \,,\qquad
b_{\mu\nu}\equiv\delta B_{\mu\nu}\,.
\eeqn
The linearized Einstein tensor 
in terms of the covariant derivative $\bar{\nabla}_\mu$ compatible with the
background metric $\bar{g}_{\mu\nu}$ is given by, 
\begin{align}\label{kinopds}
{\mathcal{E}}^{~~\rho\sigma}_{\mu\nu}m_{\rho\sigma} 
=-\tfrac{1}{2}\Big[\delta^\rho_\mu\delta^\sigma_\nu\bar{\nabla}^2
+\bar g^{\rho\sigma}\bar{\nabla}_\mu\bar{\nabla}_\nu 
&-\delta^\rho_\mu\bar{\nabla}^\sigma\bar{\nabla}_\nu
-\delta^\rho_\nu\bar{\nabla}^\sigma\bar{\nabla}_\mu \nn\\
&-\bar{g}_{\mu\nu}\bar g^{\rho\sigma}\bar{\nabla}^2 
+\bar{g}_{\mu\nu}\bar{\nabla}^\rho\bar{\nabla}^\sigma\Big]m_{\rho\sigma}\,.
\end{align}
\begin{subequations}
The masses for the spin-2 fluctuation $m_{\mu\nu}$ and the antisymmetric fluctuation $b_{\mu\nu}$ are,
\beqn
m_\mathrm{FP}^2&=&m^2\big(1+\alpha^{-2}c^{-2}\big)\Big( 3 b_1c  + 4 b_2c^2 + 3 b_3c^3 \Big)\,,\\
m_b^2&=& -\frac{m_g^2m^2}{3c^2m_B^2} \Big( 3 b_1c  + 4 b_2c^2 + 3 b_3c^3 \Big) \,.
\eeqn
\end{subequations}
We note that these two masses are related by 
$m_b^2=-\frac{\alpha^2 m_g^2}{3m_B^2(1+\alpha^2c^2)} m_\mathrm{FP}^2$.
The linearized spectrum described by \eqref{lineq} consists of one massless spin-2, 
one massive spin-2 and one massive antisymmetric field with a tachyonic mass pole
(at least for $c^2>0$).\footnote{We thank James Bonifacio for pointing this out.}
The number of propagating
degrees of freedom is therefore $2+5+3=10$. In appendix~\ref{app:ADM} we confirm
that the number of degrees of freedom is the same in the nonlinear theory.

\section{Generalization to multiple vierbeine}\label{sec:multi}

In this section we further generalize the bigravity theory with antisymmetric components to
the case of $\mathcal{N}$ dynamical vierbein fields ${(e_I)_\mu}^a$ with $I=1,\hdots,\mathcal{N}$.
We define the respective metric tensors as $(g_I)_{\mu\nu}={(e_I)_\mu}^a\eta_{ab}{(e_I)^b}_\nu$.

\subsection{General structure}

Ghost-free multi-vierbein theories contain the $\mathcal{N}$ Einstein-Hilbert kinetic terms,
\beqn
S_g=\sum_{I=1}^\mathcal{N}m_I^2\int\dd^4x \,\sqrt{g_I} \,\Big(R(g_I) -2\Lambda_I\Big)\,.
\eeqn
For $\mathcal{N}$ vierbein fields there exist $\frac12\mathcal{N}(\mathcal{N}-1)$ antisymmetric
tensor combinations of the form $\eta_{ab}{(e_I)_{[\mu}}^a(e_J)_{\nu]}{}^b$
with $I\neq J$. Since the $\mathcal{N}$ vierbeine contain $6\mathcal{N}$ Lorentz components,
only $\mathcal{N}$ of the antisymmetric tensors can be taken to be independent. 
Furthermore, the overall Lorentz
invariance of the multi-vierbein actions will render one combination unphysical.
We can thus choose $(\mathcal{N}-1)$ independent combinations to define $(\mathcal{N}-1)$
antisymmetric tensor fields. The most convenient choice of these combinations depends
on the types of couplings present in the multi-vierbein action. We will discuss 
several explicit examples below. The kinetic terms for the antisymmetric components 
in the action read,
\beqn
S_B=-  \frac1{2\cdot3!}\sum_{I=1}^\mathcal{N}\int\dd^4x~ \sqrt{g}~(H_I)_{\mu\nu\rho}(H_I)^{\mu\nu\rho}\,,
\eeqn
with $(H_I)_{\mu\nu\rho}=3 \nabla_{[\mu}(B_I)_{\nu\rho]}$ and where $\gmn$ is one metric
which has picked out of the $\mathcal{N}$ symmetric fields $(g_I)_{\mu\nu}$.
Moreover, the action will contain a potential,
\beqn
S_\mathrm{int}=\int\dd^4 x ~V(e_I)\,,
\eeqn
and thus have the total form $S=S_g+S_B+S_\mathrm{int}$.

The interactions among the vierbein fields can now have two distinct forms: They can
be \textit{pairwise} couplings~\cite{Hinterbichler:2012cn}, corresponding to multiple copies
of the bigravity case, or they can consist of \textit{determinant} vertices~\cite{Hassan:2018mcw}, 
which are genuine multi spin-2 interactions involving more than just two vierbeine in one vertex.
The pairwise couplings further split up into two categories: The \textit{center} coupling, where one vierbein
in the center interacts with all other vierbeine, and the \textit{chain} coupling, in which each 
vierbein (except for the two at the ends of the chain) interacts with exactly two neighbours.
The two distinct types of pairwise interaction graphs are displayed in Fig.~1; the left panel
of Fig.~2 shows the determinant vertex.

 \begin{figure}[h]
   \begin{center}$
   \begin{array}{cc}
   \includegraphics[width=100pt]{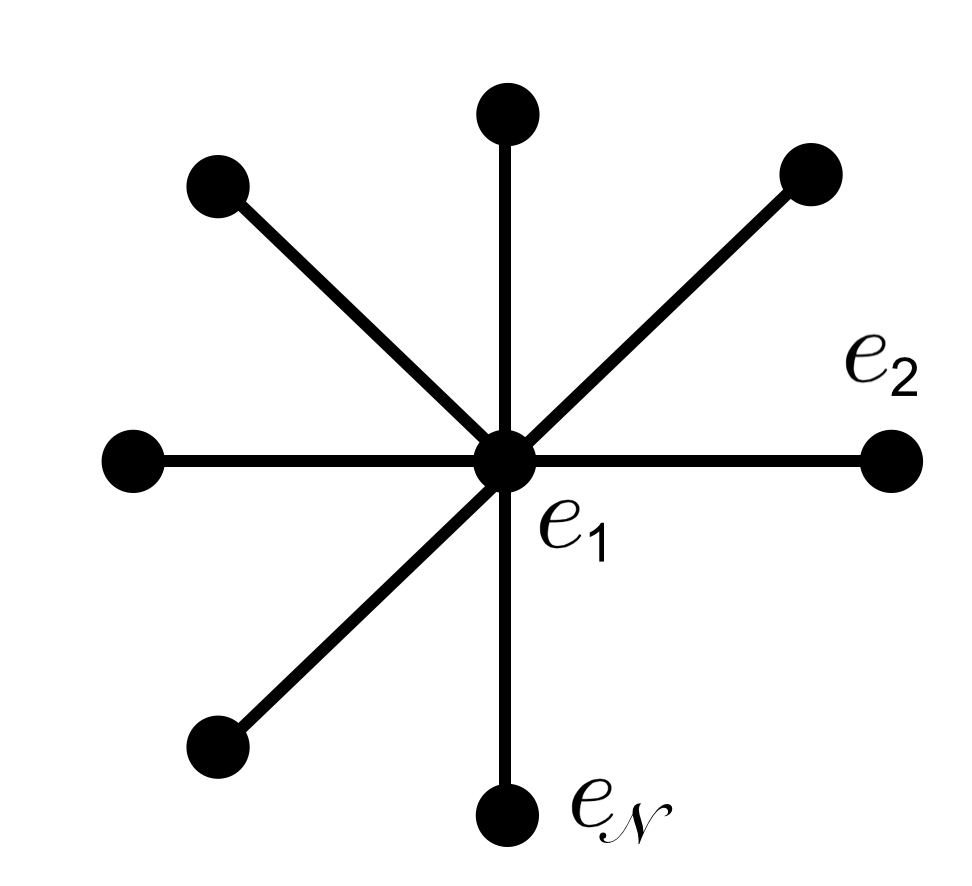}~~~~~~~~ & ~~~~~~~~
   \includegraphics[width=180pt]{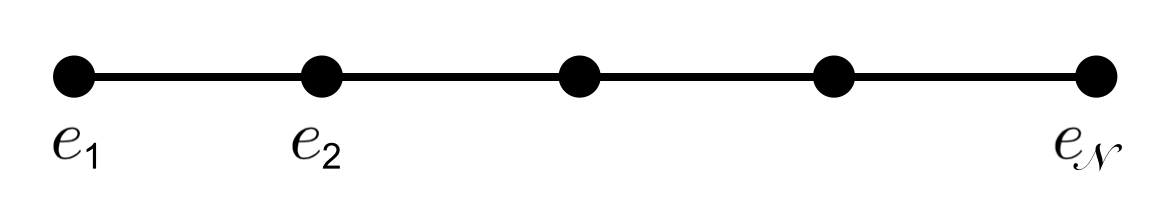}
   \end{array}$
   \caption{\textit{Left:} Center coupling of $\mathcal{N}$ vierbeine. 
   The fields $e_I$ are represented by black dots, the black lines stand for pairwise interactions 
   of the form~\eqref{defpot3}.
   \textit{Right:} Chain coupling of $\mathcal{N}$ vierbeine.}
   \end{center}
   \end{figure}

The most general vierbein theory contains all these couplings. The only two restrictions 
are that the graph of vierbein interactions can never be closed into a loop and that 
no two vierbeine can share more than one determinant vertex. An example for such a graph is
displayed in the right panel of Fig.~2.
In the following we discuss giving dynamics to the antisymmetric components for the 
different types of couplings one by one.

\begin{figure}[h]\label{fig2}
   \begin{center}$
   \begin{array}{cc}
   \includegraphics[width=90pt]{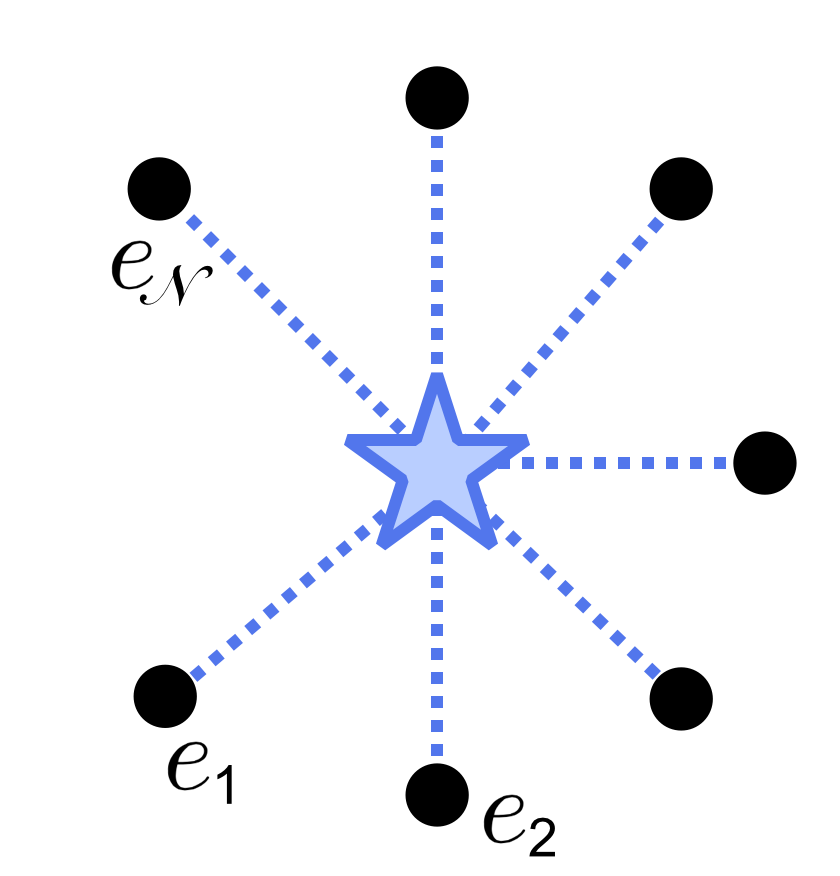}~~~~~~~~ & ~~~~~~~~
   \includegraphics[width=120pt]{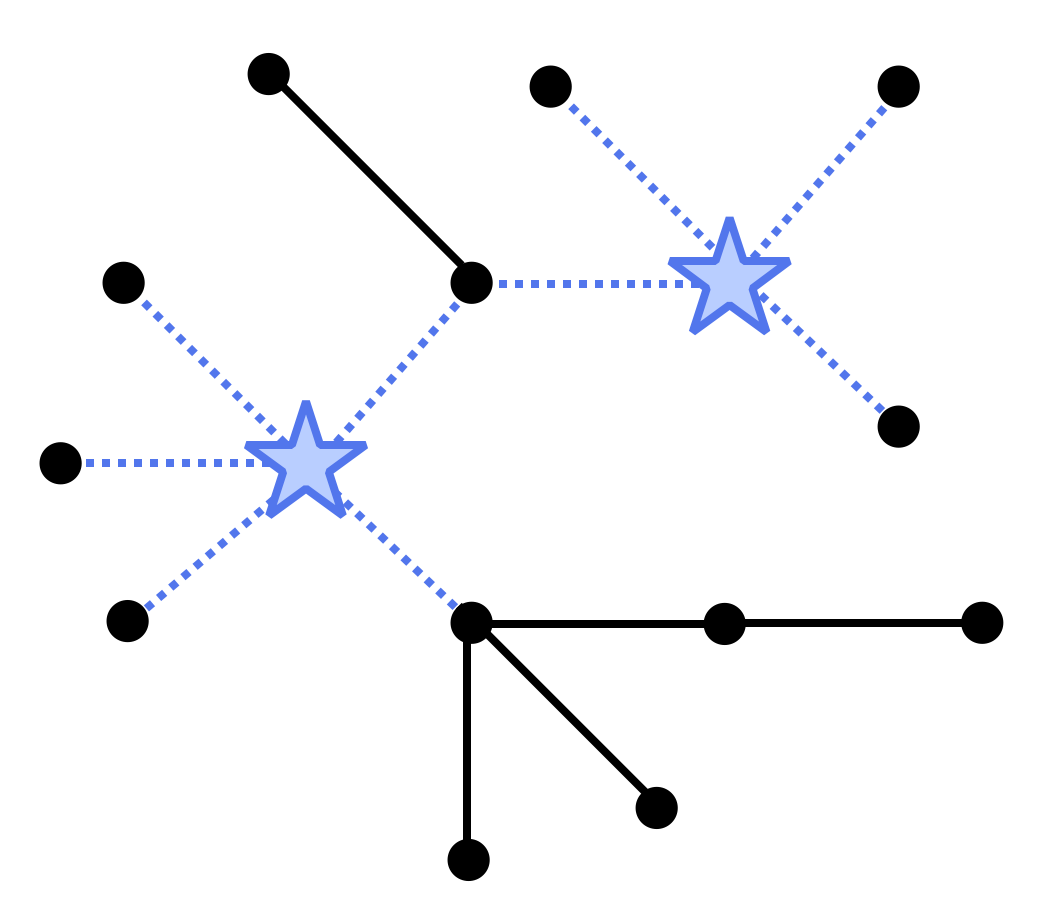}
   \end{array}$
   \caption{\textit{Left:} Determinant coupling of $\mathcal{N}$ vierbeine. 
   The fields $e_I$ are represented by black dots, the blue dashed lines ending in the star 
   stand for the multi-vierbein vertex.
   \textit{Right:} A general graph of ghost-free multi-vierbein interactions involving all types
   of couplings.}
   \end{center}
   \end{figure}

\subsection{Pairwise couplings}

\subsubsection{Center coupling}
Denoting the three interaction terms in \eqref{defpot3} by $V(e, \te\, ; b_n)$, the 
center coupling is a sum of pairwise vierbein interactions of the form,
\beqn
V_\mathrm{center}=\sum_{I=2}^\mathcal{N} V(e_I, e_1; b_n^I)\,.
\eeqn
The vierbein ${(e_1)_\mu}^a$ is in the center of the interaction graph and couples 
directly to all other vierbeine, which do not have any direct interactions among
themselves, cf.~the left panel of Fig.~1.
In this case, we define the set of $(\mathcal{N}-1)$ antisymmetric tensors as,
\beqn
(B_I)_{\mu\nu}\equiv {(e_I)_\mu}^a\eta_{ab}{(e_1)^b}_\nu -{(e_1)_\mu}^a\eta_{ab}{(e_I)^b}_\nu\,,
\eeqn
for $I=2,\hdots,\mathcal{N}$. These fields are given $(\mathcal{N}-1)$ kinetic terms
which can be covariantized independently using any of the metrics $(g_I)_{\mu\nu}$.
A straightforward generalization of the ADM analysis in appendix~\ref{app:ADM} then verifies
that the number of propagating degrees of freedom in this theory is 
$2+5(\mathcal{N}-1)+3(\mathcal{N}-1)$, corresponding to one massless spin-2, $(\mathcal{N}-1)$ massive
spin-2 and $(\mathcal{N}-1)$ massive antisymmetric tensor fields.

\subsubsection{Chain coupling}
The chain coupling is also a sum of pairwise vierbein interactions,
\beqn
V_\mathrm{chain}=\sum_{I=2}^\mathcal{N} V(e_I, e_{I-1}; b_n^I)\,.
\eeqn
The vierbeine ${(e_1)_\mu}^a$ and ${(e_\mathcal{N})_\mu}^a$ sit on the ends of the
chain, in which any other vierbein interacts with its two neighbours only.
In this case, we define,
\beqn 
(B_I)_{\mu\nu}\equiv {(e_I)_\mu}^a\eta_{ab}{(e_{I-1})^b}_\nu -{(e_{I-1})_\mu}^a\eta_{ab}{(e_{I})^b}_\nu\,,
\eeqn
for $I=2,\hdots,\mathcal{N}$, and give kinetic terms to these $(\mathcal{N}-1)$ antisymmetric
tensors. Again each of these kinetic terms can be covariantized with any of the metrics $(g_I)_{\mu\nu}$. 
As in the case of the center coupling, the propagating degrees of freedom are
one massless spin-2, $(\mathcal{N}-1)$ massive
spin-2 and $(\mathcal{N}-1)$ massive antisymmetric tensor fields.

\subsection{Determinant vertex}

The genuine multiple vierbein interactions are of the following form,
\beqn
V_\mathrm{det}=\det\Big(\sum_{I=1}^\mathcal{N} e_I\Big)\,.
\eeqn
In contract to the pairwise interactions, in this coupling each vierbein interacts with all other 
$(\mathcal{N}-1)$ fields. We can now pick any vierbein, for instance ${(e_1)_\mu}^a$ and define 
the $(\mathcal{N}-1)$ independent antisymmetric tensors as,
\beqn
(B_I)_{\mu\nu}\equiv {(e_I)_\mu}^a\eta_{ab}{(e_1)^b}_\nu -{(e_1)_\mu}^a\eta_{ab}{(e_I)^b}_\nu\,.
\eeqn
for $I=2,\hdots,\mathcal{N}$. Their covariant kinetic terms can again be written using
any of the metrics $(g_I)_{\mu\nu}$.
Since the determinant vertex is also of a totally antisymmetric
structure, its 3+1 form will be a generalization of eq.~(\ref{massadm}). The ADM analysis therefore
generalizes exactly as in the case of pairwise interactions and the degrees of freedom are again
one massless spin-2, $(\mathcal{N}-1)$ massive
spin-2 and $(\mathcal{N}-1)$ massive antisymmetric tensor fields.

\section{Discussion}

We have generalized the massive gravity theory with dynamical antisymmetric components proposed 
in Ref.~\cite{Markou:2018wxo} to the case with a dynamical reference frame $\te^a_{~\mu}$. 
The difference of the model with fixed reference frame and the fully dynamical setup
is similar to the difference of massive gravity with a fixed fiducial metric and bimetric theory
with two dynamical tensor fields. In particular, the theory proposed in this work
is both local Lorentz and diffeomorphism invariant. The setup with fixed reference
vierbein can be obtained from the fully dynamical theory by taking the limit $m_f\rightarrow\infty$,
while keeping all other parameters fixed.

Maximally symmetric background solutions (i.e.~solutions that are invariant under the 
isometry groups ISO(3,1), SO(4,1) or SO(4,2)) require the vanishing of the antisymmetric
components, $\bmn=0$. As we saw, their fluctuations are tachyonic and hence the corresponding
vacua are unstable. It would be interesting to see whether this is also the case for 
other physically relevant solutions, such as spherically symmetric or homogeneous 
and isotropic backgrounds. In any case, the fluctuations of the massive antisymmetric 
tensor will introduce nontrivial effects into the perturbation theory around such 
backgrounds. Whether the nonlinear Hamiltonian is bounded from below is an open question.

Matter can be coupled to the theory in at least three different ways without exciting 
additional degrees of freedom in the gravitational sector:
\begin{itemize}
\item[(i)] through a minimal coupling to the vierbein $e^a_{~\mu}$,
\item[(ii)] through a minimal coupling to the vierbein $\te^a_{~\mu}$,
\item[(iii)] through a minimal coupling to a linear combination $e^a_{~\mu}+a\te^a_{~\mu}$ 
with an arbitrary coefficient~$a$.
\end{itemize}
All these couplings will be linear in the lapse and shifts functions of the two
vierbeine and therefore not destroy the constraint structure discussed in appendix~\ref{app:ADM}.

Interestingly, option (iii) for the matter coupling (which was suggested for bimetric theory in 
Ref.~\cite{Noller:2014sta, Hinterbichler:2015yaa}) also opens up the possibility of defining
the gravitational theory in a more symmetric way. Instead of coupling the
kinetic term for the $\bmn$ field to the metric
$\gmn$ of the vierbein $e^a_{~\mu}$, we can couple it to the metric built from the linear 
combination of vierbeine,
$G_{\mu\nu}=(e^a_{~\mu}+a\te^a_{~\mu})\eta_{ab}(e^b_{~\nu}+a\te^b_{~\nu})$.
The analysis in appendix~\ref{app:ADM} can be applied to this case with only minor modifications,
which implies that the number of propagating degrees of freedom is again the same.
Hence, we obtain another ghost-free action by replacing the metric $\gmn$ in the kinetic term for $\bmn$
in \eqref{bgbaction} by the metric $G_{\mu\nu}$.

Since the massive $\bmn$ field is dual to a massive vector in $D=4$, it 
would be interesting to see whether there exists a dual formulation of our setup. 
This would deliver an equivalent action, possibly formulated in terms of the Lorentz 
invariant components of the vierbeine (i.e.~the corresponding metric tensors $\gmn$ and
$\fmn$) and a massive vector field $A_\mu$. The dualization of the vierbein potential may
thus produce new types of interactions for massless and massive spin-2 with massive vector fields, 
possibly relating our work to generalized Proca theories~\cite{Tasinato:2014eka, Heisenberg:2014rta}.

General relativity (GR) and its interpretation in terms of Riemannian geometry are
a prime example of the interplay between geometric structures and fundamental physics. 
Understanding the underlying geometry of any theory which includes gravity is thus crucial. For example, 
the geometry of string theory and its web of dualities gives rise to interesting new mathematical structures 
such as the extended space of Double Field Theory~\cite{Siegel:1993xq,Siegel:1993th, Hull:2009mi} which 
is related to generalized geometry~\cite{Gualtieri:2003dx, Hitchin:2004ut} 
and Born geometry~\cite{Freidel:2017yuv, Freidel:2018tkj}. Interestingly,
all these setups with intimate relation to quantum gravity contain an antisymmetric 
structure in addition to the metric.
Another example is Hermitian gravity~\cite{Chamseddine:2000zu, Chamseddine:2012gh, Chamseddine:2010rv} 
which proposes a Hermitian geometry for an extension of GR and 
also includes an antisymmetric tensor field. 
The underlying geometric structures of massive spin-2 theories still need to be constructed 
and understood in detail. The extension by an antisymmetric field as presented in this 
work may provide a first step in this direction.

\vspace{3pt}

\paragraph{Acknowledgements.} We are grateful to James Bonifacio for very valuable comments on the draft. 
This work is supported by a grant from the Max-Planck-Society.

\appendix

\section{ADM analysis}\label{app:ADM}

This appendix contains an ADM constraint analysis and a
degree of freedom counting in ADM variables for the vierbeine~\cite{Arnowitt:1962hi}.

\subsection{The 3+1 parametrization}

We parameterize the general vierbein $e^a_{~\mu}$ as a Lorentz transformation of a
gauge-fixed vierbein~$E^a_{~\mu}$,
\beqn\label{vbADM}
e^a_{~\mu}&=&\Lambda^a_{~b}E^b_{~\mu}
=\begin{pmatrix}
\Gamma & \gamma v_\beta \\
\Gamma v^\alpha & \mathcal{V}^\alpha_{~\beta}
\end{pmatrix}
\begin{pmatrix}
N & 0\\
E^\beta_{~j}N^j & E^\beta_{~i}
\end{pmatrix}\,,
\eeqn
where
\beqn
\Gamma&\equiv& \frac{1}{\sqrt{1-v^\alpha v_\alpha}}\,,\qquad 
\mathcal{V}^\alpha_{~\beta}~\equiv ~\delta^\alpha_{~\beta}+\frac{\Gamma^2}{1+\Gamma}v^\alpha v_{\beta}\,.
\eeqn
Here $\alpha,\beta=1,2,3$ are spatial Lorentz and $i,j=1,2,3$ are spatial coordinate indices.
Moreover, the Lorentz rotations sit entirely in $E^\beta_{~i}$, such that we can write
\beqn
E^\beta_{~i}=R^\beta_{~\alpha}\bar{E}^\alpha_{~i}\,,
\eeqn
for some gauge-fixed $\bar{E}^\alpha_{~i}$ with 6 independent components and $R^\mathrm{T}=R^{-1}$.
The second vierbein is parameterized as,
\beqn\label{ADMte}
\te^a_{~\mu} =
\begin{pmatrix}
L & 0\\
\tE^\alpha_{~j}L^j & \tE^\alpha_{~i}
\end{pmatrix} \, ,
\eeqn
which does not require its own set of Lorentz parameters. They can be shifted into $e^a_{~\mu}$ 
because the action is invariant under diagonal local Lorentz transformations. Thus
${\tE}^\alpha_{~i}$ has only 6 independent components.

The ADM expressions for the vierbeine induces the 3+1 split for the antisymmetric tensor defined
by $\bmn=2\eta_{ab}\te^a_{~[\mu}e^b_{~\nu]}$. The result is,
\begin{subequations}
\beqn\label{badm}
B_{0i } &=& 
=   -\Gamma v_\alpha(L E^\alpha_{~i}  + N \tE^\alpha_{~i}) 
+  \tE^\alpha_{~k} \mathcal{V}_{\alpha\beta} E^\beta_{~l} (L^k\delta^l_i - \delta^k_iN^l) \,,\\
B_{ij} &=& 
= \mathcal{V}_{\alpha\beta} \tE^\alpha_{~[i} E^\beta_{~j]}
= \mathcal{V}_{\alpha\beta} R^{\beta}_{~\gamma}\tE^\alpha_{~[i}\bar{E}^\gamma_{~j]} \,.
\eeqn
\end{subequations}
In particular, $B_{0i}$ is linear in the lapses and shifts while $B_{ij}$ is independent of them.

\subsection{Kinetic terms}

The 3+1 splits for the Einstein-Hilbert terms written in terms of 
the vierbeine are of the form~\cite{Deser:1976ay} (see also~\cite{Peldan:1993hi}),
\begin{subequations}\label{EHadm}
\beqn
\mathcal{L}_e &=& \bar{\Pi}_\alpha^{~i}\dot{\bar{E}}^\alpha_{~i} - N\mathcal{C}^{(e)} - N^i\mathcal{C}^{(e)}_i\,, 
\\
\mathcal{L}_{\te} &=& \tilde{\Pi}_\alpha^{~i}\dot{\tE}^\alpha_{~i} - L\mathcal{C}^{(\te)} - L^i\mathcal{C}^{(\te)}_i 
\eeqn
\end{subequations}
where $\bar{\Pi}_\alpha^{~i}$ and $\tilde{\Pi}_\alpha^{~i}$ are the canonical momenta 
conjugate to the spatial vierbein components ${\bar{E}}^\alpha_{~i}$ and ${\tE}^\alpha_{~i}$, respectively.
The kinetic terms for $\gmn$ and $\fmn$ are linear in $N$, $N^i$, $L$ and $L^i$ 
since the constraint contributions $\mathcal{C}^{(e)}$, $\mathcal{C}^{(e)}_i $, $\mathcal{C}^{(\te)}$
and $\mathcal{C}^{(\te)}_i $ do not depend on the lapse and shift functions. 
The precise form of the constraints will not be needed in the following.


The antisymmetric tensor is split into its components $B_{0i}$ and $B_{ij}$,
for which we insert the expressions in \eqref{badm}.
As was shown in Ref.~\cite{Markou:2018wxo}, the kinetic term for $\bmn$ 
then possesses the following form,
\beqn
\mathcal{L}_B
=
\Pi^{mn}\dot{B}_{mn}  &-& \frac{N}{\sqrt{\gamma}}\Pi^{mn}\Pi_{mn}  
- 3N^{[k}\Pi^{ij]}\partial_k(\tE^\alpha_{~i}\mathcal{V}_{\alpha\beta} E^\beta_{~j}) \nn\\
&+& \partial_i\Pi^{ij}  \big[  -\Gamma v_\alpha(L E^\alpha_{~j}  + N \tE^\alpha_{~j})
+  \tE^\alpha_{~k} \mathcal{V}_{\alpha\beta} E^\beta_{~l} (L^k\delta^l_j - \delta^k_jN^l)  \big] \nn\\
&-&\frac14 N\sqrt{\gamma} \partial_k(\tE^\alpha_{~i}\mathcal{V}_{\alpha\beta} 
 E^\beta_{~j})\partial_l(\tE^\alpha_{~p}\mathcal{V}_{\alpha\beta} E^\beta_{~q}) \Xi^{kl,[ij],[pq]} \,.
\eeqn
where $\Pi^{mn}$ is the canonical momentum conjugate to ${B}_{mn}$ and
$
\Xi^{kl,ij,pq} \equiv \gamma^{lk}\gamma^{pi}\gamma^{qj} 
+ \gamma^{lj}\gamma^{pk}\gamma^{qi} + \gamma^{li}\gamma^{pj}\gamma^{qk}
$.
This can be rewritten as,
\beqn\label{bkinadm}
\mathcal{L}_B
&=&
\Pi^{mn}\dot{B}_{mn}  
-N\mathcal{C}^{(B)}-L\tilde{\mathcal{C}}^{(B)}-N^i\mathcal{C}^{(B)}_i-L^i\tilde{\mathcal{C}}^{(B)}_i \,,
\eeqn
where none of the constraint contributions 
$\mathcal{C}^{(B)}$, $\mathcal{C}^{(B)}_i$, $\tilde{\mathcal{C}}^{(B)}$
and $\tilde{\mathcal{C}}^{(B)}_i$ depend on the lapses and shifts.
As we argued in Ref.~\cite{Markou:2018wxo}, a field redefinition can relate the 
3 dynamical components ${B}_{ij}$ to the 3 Lorentz rotation parameters in $R^\alpha_{~\beta}$.

\subsection{Interaction potential}
As was first shown in  Ref.~\cite{Hinterbichler:2012cn} and reviewed in detail in 
Ref.~\cite{Markou:2018wxo}, the antisymmetric structure of the potential $V$ ensures 
its linearity in the $e^a_{~0}$ and $\te^a_{~0}$ components of the vierbeine.
This in turn implies that the potential is linear
in the lapse and shift functions of both $e^a_{~\mu}$ and $\te^a_{~\mu}$. Hence we can write,
\beqn\label{massadm}
V=N\mathcal{C}^{(V)}+N^i\mathcal{C}^{(V)}_i+L\tilde{\mathcal{C}}^{(V)}+L^i\tilde{\mathcal{C}}^{(V)}_i
\,,
\eeqn
where $\mathcal{C}^{(V)}$, $\mathcal{C}^{(V)}_i$, $\tilde{\mathcal{C}}^{(V)}$ and 
$\tilde{\mathcal{C}}^{(V)}_i$ are functions of the remaining ADM variables alone.

\subsection{Full action}

Putting together the results for the two kinetic terms and the mass potential, 
the whole action assumes the form,
\beqn
S=\int\dd^4x\Big(\bar{\Pi}_\alpha^{~j}\dot{\bar{E}}^\alpha_{~j}
+\tilde{\Pi}_\alpha^{~j}\dot{\varphi}^\alpha_{~j}
+\Pi^{ij}\dot{B}_{ij} 
-N\mathcal{C} -N^i\mathcal{C}_i
-L\tilde{\mathcal{C}} -L^i\tilde{\mathcal{C}}_i\Big)\,,
\eeqn
where the $\mathcal{C}$ and $\mathcal{C}_i$, $\tilde{\mathcal{C}}$ and $\tilde{\mathcal{C}}_i$
contain the contributions from \eqref{EHadm}, \eqref{massadm} and \eqref{bkinadm}
which do not depend on $N$, $N^i$, $L$ and $L^i$.

The equations for $v^\alpha$ can now be solved for the components of one of the shift vectors,
say $L_i$. This gives a solution for $L_i$ which is linear in $N$, $L$ and $N^i$. 
The $L_i$ equations can be solved for $v^\alpha$ and imply $\tilde{\mathcal{C}}_i=0$. 
Thus, after solving the constraints
for $L_i$ and $v^\alpha$, the action is of the form
\beqn
S=\int\dd^4x\Big(\bar{\Pi}_\alpha^{~j}\dot{\bar{E}}^\alpha_{~j}
+\tilde{\Pi}_\alpha^{~j}\dot{\varphi}^\alpha_{~j}
+\Pi^{ij}\dot{B}_{ij} 
-N\mathcal{C} -N^i\mathcal{C}_i-L\tilde{\mathcal{C}}\Big)\,,
\eeqn
where the remaining constraints $\mathcal{C}$, $\mathcal{C}_i$ and $\tilde{\mathcal{C}}$ are
functions of ${\bar{E}}^\alpha_{~j}$, ${\varphi}^\alpha_{~j}$, ${B}_{ij}$ and their canonical momenta.

The equations for $N$, $N^i$ and $L$ impose 5 constraints on the dynamical variables
${\bar{E}}^\alpha_{~j}$, ${\varphi}^\alpha_{~j}$, ${B}_{ij}$ (or $R^\alpha_{~\beta}$). The number of
propagating degrees of freedom is thus expected to be $6+6+3-5=10$, corresponding to a massless spin-2 (2), 
a massive spin-2 (5) and a massive $\bmn$ field (3).

The shift constraint 
$\mathcal{C}_i$ together with a combination of the lapse constraints $\mathcal{C}$ and $\tilde{\mathcal{C}}$ 
will generate the diagonal diffeomorphism symmetry of the action. 
The remaining combination of lapse constraints only removes a full degree of freedom if it
gives rise to a secondary constraint. We do not explicitly verify these expected features here.
For pure bimetric theory, they were confirmed in Ref.~\cite{Hassan:2011ea, Hassan:2018mbl}.


\end{document}